\pdfoutput=1
\documentclass[a4paper,american,floatfix,pdftex,superscriptaddress,twoside,%
aps,%
reprint%,
]{revtex4-2}%
\usepackage[T1]{fontenc}
\usepackage{multirow}
\usepackage{graphicx}%
\usepackage[utf8]{inputenc}
\usepackage{hyperref, hypernat}
\usepackage[ams,todos]{CMImacros}
\usepackage[boxed]{algorithm2e}
\graphicspath{{./figures/}}
\usepackage{tabularx}
\usepackage{longtable}
\usepackage{booktabs}
\usepackage{listings}
\usepackage{xcolor} % for custom colors (optional)

\usepackage[frozencache,cachedir=.]{minted} 

\newcommand{\cfeldesy}{\affiliation{Center for Free-Electron Laser Science CFEL, Deutsches
      Elektronen-Synchrotron DESY, Notkestr. 85, 22607 Hamburg, Germany}}%

\newcommand{\ayemail}{\email[Email:~]{andrey.yachmenev@robochimps.com}}%
\newcommand{\chimpsweb}{\homepage[\\ URL:~]{https://github.com/robochimps}}%

\SetAlCapSkip{\abovecaptionskip}

\begin{document}

\title{Taylor-mode automatic differentiation for constructing molecular rovibrational Hamiltonian operators}%
\author{Andrey Yachmenev}\ayemail\chimpsweb\noaffiliation%
\author{Emil Vogt}\cfeldesy % 
\author{Álvaro Fernández Corral}\cfeldesy %
\author{Yahya Saleh}%\cfeldesy\uhhmaths %

\date{\today}

\begin{abstract}
We present an automated framework for constructing Taylor series expansions of rovibrational kinetic and potential energy operators for arbitrary molecules, internal coordinate systems, and molecular frame embedding conditions.
Expressing operators in a sum-of-products form allows for computationally efficient evaluations of matrix elements in product basis sets.
Our approach uses automatic differentiation tools from the Python machine learning ecosystem, particularly the JAX library, to efficiently and accurately generate high-order Taylor expansions of rovibrational operators.
The implementation is available at \url{https://github.com/robochimps/vibrojet}.
\end{abstract}

\maketitle

\section{Introduction}

Accurate calculations of molecular rovibrational spectra, especially for floppy molecules, weakly bound complexes, and highly excited states in general, are often computationally demanding.
These calculations benefit significantly from the use of tailored coordinates that effectively capture the essential rovibrational motions specific to each molecule~\cite{Oenen_JCP160_2024, Schneider_JCP161_2024, Saleh_JCTC21_2025, Vogt_arXiv2502.15750_2025}.
While several advanced generalized methodologies exist to perform such calculations~\cite{Yurchenko_JMolSpec245_2007, Matyus_JCP130_2009, Wang_JCP130_2009}, applying them to a new molecule typically requires more than simply providing a potential energy surface (PES).
It also involves defining an appropriate kinetic energy operator (KEO) based on the chosen internal coordinates and molecular frame embedding conditions.
This setup process demands considerable expertise in both the theoretical foundations and practical aspects of implementation.

The challenge of choosing optimal internal coordinates and frame embeddings has been the subject of ongoing research, with most recent advances focused on variational optimization of coordinate systems~\cite{Zhang_JCP161_2024, Saleh_JCTC21_2025, Vogt_arXiv2502.15750_2025}.
However, even after suitable coordinates are defined, constructing the PES and KEO in those coordinates, and efficiently evaluating their matrix elements in a chosen basis set remains nontrivial.
The matrix elements are typically computed using multidimensional numerical integration methods, such as Gaussian quadrature, often combined with Smolyak sparse grids to mitigate the curse of dimensionality~\cite{Avila_JCP139_2013}.
Related discrete variable representation methods also build on Gaussian quadrature to define an orthonormal basis that simplifies the numerical integration~\cite{Light_AdvChemPhys_2000, Tennyson_ComPhysComm163_2004, Matyus_JCP130_2009, Wang_JCP130_2009}.
An alternative approach, which is the focus of this study, involves expressing the rovibrational operators as sums of products of univariate functions, which significantly reduces the complexity of multidimensional integration. For example, when a direct product basis of univariate functions is used, multidimensional integrals reduce to a sum of products of one-dimensional integrals.

Molecular PESs and dipole moment surfaces are commonly represented in a
sum-of-products form by fitting analytic functions to electronic structure data computed on a
grid of molecular geometries.
For the KEO, one established
approach involves using symbolic algebra to derive an analytical expression, followed by manual restructuring into a sum-of-products form and hand-coded implementation~\cite{Yurchenko_JCP153_2020}.
This process is tedious and highly system specific, it must be repeated for every new molecule type, coordinate system, and frame embedding condition.
Moreover, the embedding must be analytically expressible, which often excludes commonly used schemes like the Eckart frame due to their algebraic complexity.
The choice of embedding is particularly important when rotational motion is considered, as it affects the separability of vibrational and rotational degrees of freedom.
It influences not only the basis set convergence of rovibrational energies but also the accuracy of vibrational transition intensities, even when rotational degrees of freedom are not explicitly included in the model.

Certain choices of vibrational coordinates and embedding conditions, such as polyspherical coordinates, enable derivation of the rovibrational KEO directly in the sum-of-products form~\cite{Mladenovic_JCP112_2000,Wang_JCP113_2000,Gatti_JCP114_2001,Schwenke_JCP118_2003}.
This avoids complex symbolic algebra and facilitates the construction of the KEO from molecular fragments.

A more general and flexible alternative is to construct the rovibrational operators in a sum-of-products form \emph{via} Taylor series expansions~\cite{Yurchenko_JMolSpec245_2007,Yachmenev_JCP143_2015} or least-squares fitting techniques~\cite{Pelez_JCP138_2013, Ziegler_JCP144_2016}.
In the Taylor expansion approach, derivatives can be computed using finite-difference methods~\cite{Ovsyannikov_OptSpec107_2009} or automatic differentiation (AD), particularly \emph{via} forward-mode propagation of Taylor series~\cite{Yachmenev_JCP143_2015}.
These methods are well-suited for handling complex coordinate transformations and frame embeddings, including those defined implicitly by nonlinear equations (e.g., Eckart conditions), enabling fully automated construction of KEOs without manual symbolic manipulation.

In this work, we introduce \texttt{Vibrojet}, a general Python-based framework for constructing Taylor series expansions of molecular rovibrational operators using modern AD tools.
Specifically, we employ the \texttt{JAX} library and its \texttt{jet} module, which supports efficient Taylor-mode differentiation.
We further employ an efficient approach for computing high-order partial derivatives by propagating families of univariate Taylor series~\cite{Griewank_MathComp69_2000, Walther_thesis_2012}.
Compared to our previous Fortran-based implementation~\cite{Yachmenev_JCP143_2015}, the new Python framework offers several advantages, including easier coding for setting up new molecules and coordinate systems, integration with a broader scientific computing ecosystem, and more accessible platform-specific optimization and parallelization.

This manuscript is organized as follows.
In \autoref{sec:numeric_keo}, we present a general numerical approach for constructing the KEOs of molecules.
The Taylor-mode AD approach is described in \autoref{sec:taylor} and details of its implementation in \autoref{sec:implementation}.
Finally, \autoref{sec:examples} provides illustrative examples of Taylor series
expansion for the PES and KEO of selected molecules and demonstrates their use in
variational calculations of vibrational energy levels.

\section{Numerical procedure for KEO}\label{sec:numeric_keo}

The general form of the rovibrational KEO is given by
\begin{align*}
\hat{T} = \sum_{\lambda,\mu=1}^{M+6}p_\lambda^\dagger G_{\lambda\mu}(\boldsymbol{\xi})p_\mu + U(\boldsymbol{\xi}),
\end{align*}
where $p_{\lambda=1..M+6}=\{-i\hbar\partial/\partial\xi_1$, $...$, $-i\hbar\partial/\partial\xi_M$, $\hat{J}_x$, $\hat{J}_y$, $\hat{J}_z$,$-i\hbar\partial/\partial X_\text{cm}$, $-i\hbar\partial/\partial Y_\text{cm}$, $-i\hbar\partial/\partial Z_\text{cm}\}$ are the momentum operators conjugate to the $M$ vibrational coordinates $\boldsymbol{\xi}=\xi_1,\xi_2,...\xi_{M}$, three Cartesian components of the rotational angular momentum operator $\mathbf{J}$, and three Cartesian momentum operators of the overall translation.
The mass-weighted contravariant metric tensor $G_{\lambda\mu}$ and pseudopotential $U$ can be computed as~\cite{Sorensen_LargeAmplitudes_1979, Matyus_JCP130_2009}
\begin{align*}
&G_{\lambda\mu}=[\mathbf{g}^{-1}]_{\lambda\mu}, \\
&g_{\lambda\mu}=\sum_{i=1}^N\sum_{\alpha=x,y,z} m_i t_{i\alpha,\lambda} t_{i\alpha,\mu}, \\
&U =\frac{\hbar^2}{32}\sum_{\lambda,\mu=1}^{M}\left[ \frac{G_{\lambda\mu}}{\tilde{g}^2} \frac{\partial\tilde{g}}{\partial \xi_\lambda}\frac{\partial\tilde{g}}{\partial\xi_\mu} + 4\frac{\partial}{\partial \xi_\lambda}\left( \frac{G_{\lambda\mu}}{\tilde{g}} \frac{\partial\tilde{G}}{\partial\xi_\mu} \right) \right],
\end{align*}
where $\tilde{g}=\det(g_{\lambda\mu})$.
The $t_{i\alpha,\lambda}$ vectors are defined in terms of the Cartesian coordinates $r_{i\alpha}$ ($\alpha=x,y,z$) of the $i=1..N$ atoms in the molecule, as
\begin{align*}
t_{i\alpha,\lambda} &= \frac{\partial r_{i\alpha}}{\partial\xi_\lambda},~~~(\lambda = 1..M), \\
t_{i\alpha,M+\beta} &= \sum_{\gamma=x,y,z} \epsilon_{\alpha\beta\gamma}r_{i\gamma},~~~(\beta=x,y,z), \\
  t_{i\alpha,M+3+\beta} &=\delta_{\alpha\lambda},~~~(\beta=X_\text{cm},Y_\text{cm},Z_\text{cm}),
\end{align*}
for vibrational, rotational, and translational coordinates, respectively, where $\epsilon_{\alpha\beta\gamma}$ is the three-dimensional Levi-Civita symbol ($\epsilon = +1$ for even permutations of $\alpha$, $\beta$ and $\gamma$, and $\epsilon = -1$ for odd permutations).
The pseudopotential $U$ is a scalar operator that can be combined with the PES to avoid the need for evaluating and storing additional integrals.

From the equations above, it follows that the KEO can be constructed automatically, provided a coordinate transformation function is defined that maps the internal vibrational coordinates $\xi_\lambda$ ($\lambda=1..M$) to the Cartesian coordinates $r_{i\alpha}$ of the atoms
\begin{align}\label{eq:r_i}
\mathbf{r}_i \equiv (x_i,y_i,z_i) = \mathbf{f}_i(\xi_1,\xi_2,...\xi_{M}).
\end{align}
Defining such a transformation establishes, either implicitly or explicitly, an orientation of the Cartesian $x,y,z$ axes relative to the molecule.
This orientation may itself vary with changes of the internal coordinates.
Note that coordinate transformation must be invertible across the domain of the vibrational coordinates.
% For example, when constructing Cartesian coordinates from a Z-matrix representation of valence internal coordinates, the $z$-axis is aligned along the first bond. The second axis is oriented perpendicular to the line defined by the first two atoms, and the third axis is oriented perpendicular to the plane defined by the first three atoms.\aynote{this is not a good example, since axes will not depend on internal coordinates}

High-order Taylor expansions of the coordinate mappings in \eqref{eq:r_i}, computed using AD tools, have been successfully employed in computational chemistry to accelerate geometry optimization, reaction path search, and classical trajectory integration~\cite{Rybkin_CompChem34_2013}.

Popular choices for orientation of the molecular frame $x,y,z$ axes include the principal axes system (PAS) and the Eckart frame.
The PAS choice minimizes the off-diagonal elements in the pure rotational block of the $G$-matrix, whereas the Eckart frame is designed to minimize the off-diagonal elements in the rovibrational (Coriolis) part of the $G$-matrix.
The orientation of the molecular frame can be conveniently expressed using a rotation matrix $\mathbf{d}(\boldsymbol{\xi})$, which maps reference Cartesian coordinates of atoms (user-defined or derived from a Z-matrix) into the Eckart or PAS frame
\begin{align}\label{eq:coord_map}
\mathbf{r}_i=\mathbf{d}\cdot\bar{\mathbf{r}}_i\equiv \mathbf{d}\cdot \mathbf{f}_i(\xi_1,...,\xi_{M}).
\end{align}
The conditions defining the Eckart or PAS frames can be formulated in terms of the rotation matrix $\mathbf{d}$ as follows
\begin{align}\label{eq:eckart}
  \mathbf{u}\mathbf{d}^T - \mathbf{d}\mathbf{u}^T &= \mathbf{0},~~~(\text{Eckart}) \\ \label{eq:pas}
  [\mathbf{d}\bar{\mathbf{u}}\mathbf{d}^T]_{\alpha\beta}&=0,~~~\alpha\neq \beta,~~~(\text{PAS}).
\end{align}
Here, the matrices $\mathbf{u}$ and $\bar{\mathbf{u}}$ are defined as
\begin{align*}
  u_{\alpha\beta} &= \sum_{i=1}^N m_i r_{i\alpha}^{\text{(ref)}}\bar{r}_{i\beta}, \\
  \bar{u}_{\alpha\beta} &= \sum_{i=1}^N m_i \bar{r}_{i\alpha}\bar{r}_{i\beta}.
\end{align*}
The Eckart conditions are defined with respect to a reference geometry, typically chosen as the equilibrium geometry $r_{i\alpha}^{\text{(ref)}}=r_{i\alpha}^{\text{(eq)}}$, around which the rovibrational coupling elements of the $G$-matrix are minimized.

The frame embedding conditions can be incorporated directly into the coordinate transformation function $\mathbf{f}_i$ in \eqref{eq:coord_map}, or handled externally through generalized routines.
We briefly outline a numerical solution of the Eckart equations, originally presented in \cite{Yachmenev_JCP143_2015}.
Several alternative approaches have been proposed based on eigenvalue problems~\cite{Dymarsky_JCP122_2005, Krasnoshchekov_JCP140_2014}, similar to the diagonalization of the moment of inertia tensor used in constructing the PAS frame.
However, we found that methods based on eigenvalue decomposition can encounter difficulties when computing derivatives required for constructing the Taylor series expansions of the KEO (see \autoref{sec:taylor}).
These issues are particularly pronounced in highly symmetric molecules such as methane, where degenerate eigenvalues arise~\cite{vibrojet_eckart_quaternion_issue} - not to be confused with eigenvalues of the Hamiltonian.
In such cases, the derivatives of eigenvectors are not well defined, making differentiation of  the eigenvalue-based solutions fundamentally problematic.

The present solution is based on parametrization of the rotation matrix $\mathbf{d}$ in \eqref{eq:coord_map} using a $3\times 3$ skew-symmetric matrix $\boldsymbol{\kappa}$, such that
\begin{align}\label{eq:dmat_expkappa}
  \mathbf{d} &= e^{-\boldsymbol{\kappa}}, \nonumber \\
  \boldsymbol{\kappa}^T &= -\boldsymbol{\kappa}.
\end{align}
The exponential parametrization ensures that the orthogonality of $\mathbf{d}$ is preserved for all values of the three independent elements $\kappa_{xy}$, $\kappa_{xz}$, and $\kappa_{yz}$.
These elements are determined by solving the Eckart equations in \eqref{eq:eckart} rewritten in terms of $\boldsymbol{\kappa}$.
Substituting \eqref{eq:dmat_expkappa} into the Eckart conditions and rearranging terms, we obtain the following linear system for $\boldsymbol{\kappa}$
\begin{align}\label{eq:eckart_solution}
\left(\begin{array}{ccc}
  u_{xx} + u_{yy} & u_{yz} & -u_{xz} \\
  u_{zy} & u_{xx}+u_{zz} & u_{xy} \\
  -u_{zx} & u_{yx} & u_{yy}+u_{zz} 
\end{array}\right) \cdot
\left(\begin{array}{c}
  \kappa_{xy} \\
  \kappa_{xz} \\
  \kappa_{yz}
\end{array}\right) \\ \nonumber
=\sum_{\alpha=x,y,z}
\left(\begin{array}{c}
  \lambda_{x\alpha}u_{y\alpha}-\lambda_{y\alpha}u_{x\alpha} \\
  \lambda_{x\alpha}u_{z\alpha}-\lambda_{z\alpha}u_{x\alpha} \\
  \lambda_{y\alpha}u_{z\alpha}-\lambda_{z\alpha}u_{y\alpha}
\end{array}\right),~~~~~~~~
\end{align}
where
\begin{align*}
  \boldsymbol{\lambda} = e^{-\boldsymbol{\kappa}} + \boldsymbol{\kappa}.
\end{align*}
This system is solved iteratively, starting with the initial guess $\boldsymbol{\lambda}=\mathbf{I}$.
At each iteration, the elements of $\boldsymbol{\kappa}$ are updated, and the exponential matrix $e^{-\boldsymbol{\kappa}}$ is computed using a Taylor series expansion, Rodrigues' formula, or the Pad{\'e} approximation~\cite{Moler_SIAMReview45_2003}.

\section{Taylor polynomials}\label{sec:taylor}

Standard AD relies on the fact that a function to be differentiated can be expressed as a composition of smooth elementary functions.
For example, if a function is defined as $f(x)=g\circ h(x)$, then by chain rule, its derivative is
$f'(x)=  g' \circ h(x) \  h'(x)$.
For higher-order derivatives of such compositions, the Faà di Bruno formula generalizes the chain rule and enables recursive computation of higher-order derivatives from lower-order ones.
If the functions involved in the composition admit converging Taylor series
expansions, a convenient method for computing higher-order derivatives is to use truncated Taylor
polynomials.
Let the function $h(x)$ be represented by the Taylor expansion in $x$ around $x_0=0$
\begin{align*}
h(x) = h_0 + h_1x +\frac{1}{2}h_2x^2 + \frac{1}{3!}h_3x^3+...+\frac{1}{n!}h_nx^n.
\end{align*}
Assuming both $h$ and $g$ admit converging Taylor expansions, the composite function $f(x)=g\circ h(x)$ can also be expressed as a Taylor series around $x_0=0$
\begin{align*}
f(x) = f_0 + f_1x +\frac{1}{2}f_2x^2 + \frac{1}{3!}f_3x^3+...+\frac{1}{n!}f_nx^n,
\end{align*}
with coefficients given by
\begin{align*}
f_0 &= g(h_0) \\
f_1 &= g'(h_0)h_1  \\ \nonumber
f_2 &= g'(h_0)h_2 + g''(h_0) h_1 h_1 \\ \nonumber
f_3 & = g'(h_0)h_3 + 3 g''(h_0) h_1 h_2 + g'''(h_0)h_1h_1h_1 \\ \nonumber
&\vdots
\end{align*}
Each coefficient $f_k$ corresponds to the $k$-th derivative of the composition $g\circ h(x)$ with respect to $x$, as described by the Faà di Bruno formula. The terms $g^{(k)}(h_0)$ represent the coefficients of the Taylor series expansion of $g$ around $h_0$.

To extend AD to higher-order derivatives, it is necessary to reformulate elementary arithmetic operations, intrinsic functions (e.g., $\exp,~\sin$), and linear algebra routines (e.g., $\det,~\text{eigh}$) to operate on truncated Taylor polynomials.
This can be accomplished by systematically applying the Faà di Bruno rule for function composition and the Leibniz rule for product differentiation.

Several prior works have implemented efficient polynomial arithmetic for truncated Taylor expansions.
Notable contributions include the foundational work of Griewank and Walther~\cite{Griewank_book_2008,Griewank_MathComp69_2000, Walther_thesis_2012} and their C/C++ implementation~\cite{Griewank_ACMTansMathSoft22_1996}, the Julia implementation by Benet and Sanders~\cite{Benet_JOpenSourceSoft4_2019}, and the Python-based approach by Bettencourt, Johnson, and Duvenaud~\cite{Bettencourt_jaxjet_2019} implemented in the \texttt{JAX} library~\cite{jax2018github}.
In our own earlier work, Taylor polynomial arithmetic in Fortran~90 was implemented as part of the rovibrational code TROVE~\cite{Yachmenev_JCP143_2015}.

In this work, we use the \texttt{jax.experimental.jet} module of the \texttt{JAX} library~\cite{jax2018github}.
Although \texttt{jet} provides high-oder derivative rules for many functions and linear algebra operations, at the time of writing, it lacks support for several essential primitives. These include eigenvalue and LU decompositions, matrix determinant, matrix inverse, and matrix exponential.
Furthermore, certain arithmetic operations specific to the KEO construction, such as solutions of the molecular frame embedding equations in \eqref{eq:eckart} and \eqref{eq:pas}, can be more efficiently implemented by expressing them directly within the Taylor polynomial algebra framework.
To that end, our code \texttt{Vibrojet} extends \texttt{JAX}'s Taylor polynomial functionality by adding support for missing linear algebra primitives required for the KEO construction.
It also introduces new operations tailored to the Eckart and PAS frame embedding equations.
For eigenvalue decomposition, we utilized formulas derived in~\cite{Mach_arXiv2302.03661_2023}.

As an example, we describe the computation of high-order derivatives of the frame rotation matrix in \eqref{eq:dmat_expkappa}, which satisfies the Eckart embedding conditions.
Differentiating \eqref{eq:eckart} $l$ times, we obtain
\begin{align*}
\sum_{m=0}^l {l\choose m} \left( \mathbf{u}_m \mathbf{d}_{l-m}^T - \mathbf{d}_{l-m} \mathbf{u}_{m}^T\right)=0,
\end{align*}
where the subscript $l$ denotes the $l$-th derivative.
Rearranging terms yields a more convenient form for recursive computation of derivatives $\mathbf{d}_l$
\begin{align} \label{eq:eckart_deriv}
\mathbf{u}\boldsymbol{\kappa}_l + \boldsymbol{\kappa}_l\mathbf{u}^T &= \boldsymbol{\lambda}_l\mathbf{u}^T-\mathbf{u}\boldsymbol{\lambda}_l^T \\ \nonumber
&+\sum_{m=1}^{l}{l\choose m}\left( \mathbf{d}_{l-m} \mathbf{u}_{m}^T - \mathbf{u}_m \mathbf{d}_{l-m}^T \right),
\end{align}
with
\begin{align}\label{eq:lambda}
\boldsymbol{\lambda}_l = \mathbf{d}_l+\boldsymbol{\kappa}_l,
\end{align}
and matrix exponential
\begin{align}\label{eq:exp_kappa_taylor}
\mathbf{d}_l = \big[ e^{-\boldsymbol{\kappa}}\big]_l = \sum_{n}\frac{(-1)^n}{n!}\big[\boldsymbol{\kappa}^n\big]_l.
\end{align}
This system of equations is solved iteratively for each derivative $\boldsymbol{\kappa}_l$, starting with the initial value $\boldsymbol{\lambda}_l=0$ for $l>\mathbf{0}$ and $\boldsymbol{\lambda}_0=\mathbf{I}$.
At each iteration, the three independent elements $\kappa_l^{(xy)}$, $\kappa_l^{(xz)}$, and $\kappa_l^{(yz)}$ are determined by solving a linear system.
The corresponding linear-system matrix remains constant across all derivative orders (same as in \eqref{eq:eckart_solution}), allowing its inverse to be computed once and reused.
The right-hand side of \eqref{eq:eckart_deriv} depends on $\boldsymbol{\lambda}_l$ (also as in \eqref{eq:eckart_solution}), which is updated recursively using \eqref{eq:lambda} and \eqref{eq:exp_kappa_taylor}.
It also includes a sum over low-order derivatives $\mathbf{d}_k$ ($k=0..l-1$), which are computed in the earlier steps of the recursion.

Derivatives of matrix powers $\left[\boldsymbol{\kappa}^n\right]_l$ are computed using the Leibniz product rule
\begin{align*}
\big[\boldsymbol{\kappa}^n\big]_l = \sum_{m=0}^{l}{l\choose m}\big[\boldsymbol{\kappa}^{n-1}\big]_m\big[\boldsymbol{\kappa}\big]_{l-m}.
\end{align*}
To ensure convergence of the exponential series in \eqref{eq:exp_kappa_taylor}, we use the scaling and squaring technique~\cite{Moler_SIAMReview45_2003}.
In practice, truncating the expansion after the 6th-order term with two to four scaling steps yields accurate results.
The solution of the Eckart equations typically converges within 6--10 iterations.
All operations are formulated as products of $3\times 3$ matrices, allowing efficient execution in Python/\texttt{JAX} using just-in-time compilation.

Finally, a note on multivariate partial derivatives.
The coefficients of multivariate Taylor polynomials can be efficiently computed by propagating univariate directional derivatives using the interpolation approach of Griewank, Utke and Walther~\cite{Griewank_MathComp69_2000} (see also Chapter 13.3 in \cite{Griewank_book_2008}).
This method is implemented in \texttt{Vibrojet} and is especially effective for Taylor expansions of rovibrational operators in large molecules, where an $N$-mode truncation scheme is commonly employed to improve convergence~\cite{Bowman_MolPhys106_2008}.
The interpolation of directional derivatives enables computation of partial derivatives along selected directions without constructing the full high-dimensional derivative tensor, resulting in significant computational savings.

\section{Implementation}\label{sec:implementation}

The mapping from internal to Cartesian coordinates, as described in \eqref{eq:coord_map}, is implemented \emph{via} a user-defined function with the following interface
\begin{align*}
x = \texttt{internal\_to\_cartesian}(q)
\end{align*}
Here, $q$ is an array of length $3N-6$ containing internal coordinate values, and $x$ is an array of shape $(N,3)$, containing the Cartesian coordinates of the $N$ atoms in the molecule.
Users can implement frame embeddings either directly within the \texttt{internal\_to\_cartesian} function or by applying predefined function decorators for common embeddings.
For example, to enforce the Eckart frame, the \texttt{eckart} decorator can be applied, which rotates the coordinate system accordingly
\begin{align*}
\texttt{eckart}(q_0, m)(\texttt{internal\_to\_cartesian})
\end{align*}
Here, $q_0$ is an array of reference (e.g., equilibrium) internal coordinates, and $m$ is an array of atomic masses.
Likewise, to shift coordinates to the centre of mass, users can apply the $\texttt{com}(m)(\texttt{internal\_to\_cartesian})$ decorator.

Below is an example of a coordinate mapping for a triatomic H$_2$O molecule using valence internal coordinates and the Eckart frame.
\begin{minted}[breaklines, breakanywhere=true, breakautoindent=true,fontsize=\small, linenos, numbersep=7pt, xleftmargin=1em]{python}
from jax import numpy as jnp
from vibrojet.eckart import eckart

# Masses of O, H, H atoms
masses = [15.9994, 1.00782505, 1.00782505]

# Equilibrium values of valence coordinates
r1, r2, alpha = 0.958, 0.958, 1.824
q0 = [r1, r2, alpha]

@eckart(q0, masses)
def valence_to_cartesian(q):
    r1, r2, a = q
    return jnp.array([
      [0.0, 0.0, 0.0],
      [r1*jnp.sin(a/2), 0.0, r1*jnp.cos(a/2)],
      [-r2*jnp.sin(a/2), 0.0, r2*jnp.cos(a/2)],
    ])

xyz0 = valence_to_cartesian(q0)
\end{minted}

Given a coordinate mapping function, the kinetic energy $G$-matrix and pseudopotential $U$ can be evaluated at specific internal coordinate values or over a grid using functions from the \texttt{keo} module: \texttt{Gmat}, \texttt{batch\_Gmat}, \texttt{pseudo}, and \texttt{batch\_pseudo}.
To compute the $G$-matrix at a single point
\begin{align*}
g = \texttt{Gmat}(q, m, \texttt{internal\_to\_cartesian})
\end{align*}
Here, $g$ is an array of shape $(3N, 3N)$ representing the full $G$-matrix.
The first $3N-6$ rows and columns correspond to vibrational coordinates, followed by three rotational and three translational coordinates.
All values are given in units of cm$^{-1}$, assuming input bond distances are in \AA\ and angles in radians.
For evaluating over a batch of points
\begin{align*}
g = \texttt{batch\_Gmat}(q, m, \texttt{internal\_to\_cartesian})
\end{align*}
In this case, $q$ is a 2D array of shape $(D, 3N-6)$, where $D$ is the number of grid points.
The output $g$ has shape $(D, 3N, 3N)$.
The functions \texttt{pseudo} and \texttt{batch\_pseudo} follow the same interface and unit convention.

Below is an example of computing the $G$-matrix and $U$ at a single point and over a grid using the coordinate mapping for the H$_2$O molecule defined above.
\begin{minted}[breaklines, breakanywhere=true, breakautoindent=true,fontsize=\small, linenos, numbersep=7pt, xleftmargin=1em]{python}
from vibrojet.keo import Gmat, batch_Gmat, pseudo, batch_pseudo

masses = [15.9994, 1.00782505, 1.00782505]
r1, r2, alpha = 0.958, 0.958, 1.824

# compute G and U at single point
q0 = [r1, r2, alpha]
g = Gmat(q0, masses, valence_to_cartesian)
u = pseudo(q0, masses, valence_to_cartesian)

# compute G and U at grid points
q = jnp.array(
    [
        [r1, r2, alpha],
        [r1 + 0.1, r2, alpha],
        [r1, r2 + 0.1, alpha],
        [r1, r2, alpha + 0.1],
    ]
)
g = batch_Gmat(q, masses, valence_to_cartesian)
u = batch_pseudo(q, masses, valence_to_cartesian)
\end{minted}

The Taylor series coefficients of a function can be computed using the \texttt{deriv\_list} function from the \texttt{taylor} module.
It takes the function to be differentiated, an expansion point, and a list of multi-indices specifying the desired derivative terms
\begin{align*}
\text{coefs} = \texttt{deriv\_list}(\text{func}, x_0, \text{deriv\_ind}, \text{if\_taylor}).
\end{align*}
The boolean parameter \texttt{if\_taylor} determines the output: if set to \texttt{True}, the function returns Taylor series coefficients; if set to \texttt{False}, it returns the corresponding partial derivatives.
For example, to compute the Taylor expansion of the $G$-matrix for H$_2$O molecule up to fourth order
\begin{minted}[breaklines, breakanywhere=true, breakautoindent=true,fontsize=\small, linenos, numbersep=7pt, xleftmargin=1em]{python}
from vibrojet.taylor import deriv_list
import itertools

# Generate the list of multi-indices specifying
# the integer exponents for each coordinate
# in the Taylor series expansion
max_order = 4  # max total expansion order
deriv_ind = [
    elem
    for elem in itertools.product(*[range(0, max_order + 1) for _ in range(len(q0))])
    if sum(elem) <= max_order
]

# Function for computing kinetic G-matrix
# for the given masses of atoms and coordinates
func = lambda x: Gmat(x, masses, valence_to_cartesian)

g_coefs = deriv_list(func, deriv_ind, q0, if_taylor=True)
\end{minted}
The output \texttt{g\_coefs} is an array where the first dimension indexes the corresponding derivative multi-indices in \texttt{deriv\_ind} and the second and third dimensions correspond to the rows and columns of the $G$-matrix.
Each entry in \texttt{g\_coefs} gives the Taylor series coefficient of a $G$-matrix element associated with a specific multi-index, evaluated at the expansion point~\texttt{q0}.

The \texttt{deriv\_list} function is not specific to the KEO and can be employed to compute Taylor expansion coefficients for any user-defined multivariate function.
For example, in can be used to expand PESs or dipole moment surfaces in the desired coordinate coordinate system.

\section{Examples}\label{sec:examples} 

We present variational calculations of the vibrational energies of formaldehyde (H$_2$CO) and ammonia (NH$_3$), using the KEO and PES represented by Taylor series expansions around a single reference configuration.
These expansions are casted into an $N$-mode representation, and the convergence of the computed vibrational energies is examined with respect to both the Taylor expansion order and the $N$-mode truncation order of both the KEO and PES.
Complete example calculations are available in the \texttt{examples} folder of \texttt{Vibrojet} repository.

For molecules like NH$_3$, which exhibit multiple minima in the PES along one or more large-amplitude coordinates, the standard approach is the Hougen-Bunker-Johns method~\cite{Hougen_JMolSpec34_1970}.
This method treats large-amplitude coordinates on a grid, while constructing Taylor expansions for the remaining quasi-rigid coordinates at each grid point.
However, it has been shown that reasonably accurate expansions of the KEO and PES for NH$_3$ can also be obtained using rectilinear coordinates, by expanding around a reference configuration corresponding to the planar equilibrium geometry~\cite{Neff_SpecActaA119_2014}.
In our NH$_3$ example, we employ this alternative approach, treating the inversion coordinate within a Taylor expansion. 
We show that the PES, kinetic $G$-matrix and pseudopotential can be efficiently expanded in terms of functions of individual internal curvilinear coordinates, such as Morse functions for stretching coordinates and trigonometric functions for bending coordinates, resulting in improved convergence of the Taylor series even for the inversion vibrational motion.

\begin{figure}
	\includegraphics[width=\linewidth]{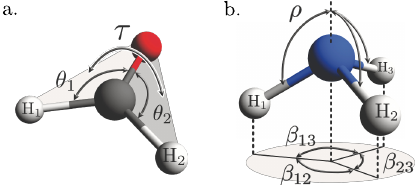} 
	\caption{Definition of the internal angular coordinates used in vibrational energy calculations for the H$_2$CO and NH$_3$ molecules.
  }
	\label{fig:nh3_h2co_coo}
\end{figure}

For H$_2$CO, the vibrational problem is solved using valence bond coordinates: $r_1$, $r_2$, $r_3$ representing the C--O, C--H$_1$, and C--H$_2$ bond lengths, respectively; two bond angles $\theta_1$ and $\theta_2$; and one dihedral angle $\tau$, as defined in~\autoref{fig:nh3_h2co_coo}.a.
The $G$-matrix, pseudopotential, and PES, originally published in \cite{Al-Refaie_MNRAS448_2015} and reimplemented in Python,
are expanded in these coordinates around the equilibrium configuration: $r_1=1.2034$~\AA, $r_2=r_3=1.1038$~\AA, $\theta_1=\theta_2=121.84^\circ$, $\tau=180^\circ$.

For NH$_3$, the vibrational problem is solved using internal coordinates: $r_1$, $r_2$, $r_3$ representing the N--H$_1$, N--H$_2$, and N--H$_3$ bond lengths, respectively; $s_4=(2\beta_{23}-\beta_{13}-\beta_{12})/\sqrt{6}$ and $s_5=(\beta_{13}-\beta_{12})/\sqrt{2}$, which are symmetry-adapted combinations of bond angles; and $\rho$, the umbrella inversion coordinate, as defined in \autoref{fig:nh3_h2co_coo}.b.
Unlike H$_2$CO, we employ two sets of transformed coordinates, $y^{(G)}$ and $y^{(V)}$, for expanding the KEO ($G$-matrix and pseudopotential) and PES, respectively
\begin{eqnarray}\label{eq:y}
\begin{array}{lll}
  y_1^{(G)} = r_1 - r_1^{(0)},~~&  y_1^{(V)}=1-e^{-a_m(r_1-r_1^{(0)})}, \\ 
  y_2^{(G)} = r_2 - r_2^{(0)},~~& y_2^{(V)}=1-e^{-a_m(r_2-r_2^{(0)})}, \\ 
  y_3^{(G)} = r_3 - r_3^{(0)},~~& y_3^{(V)}=1-e^{-a_m(r_3-r_3^{(0)})}, \\ 
  y_4^{(G)} = s_4, ~~& y_4^{(V)}=s_4,\\
  y_5^{(G)} = s_5, ~~& y_5^{(V)}=s_5,\\
  y_6^{(G)} = \cos\rho, ~~& y_6^{(V)}=\sin\rho.
\end{array}
\end{eqnarray}
The KEO expansions are constructed using the $y^{(G)}$ coordinates around the reference configuration: $r_1^{(0)}=r_2^{(0)}=r_3^{(0)}=1.0116$~\AA, $s_4=s_5=0$, and $\rho=90^\circ$ (planar configuration).
The PES, adapted from~\cite{Polyansky_JMolSpec327_2016}, is re-expanded in the $y^{(V)}$ coordinates around the same reference configuration, but with $\rho=112.1^\circ$, corresponding to one of the two equivalent minima of the PES, and using the Morse exponent $a_m=2.0$~\AA$^{-1}$.

\begin{figure}
	\includegraphics[width=\linewidth]{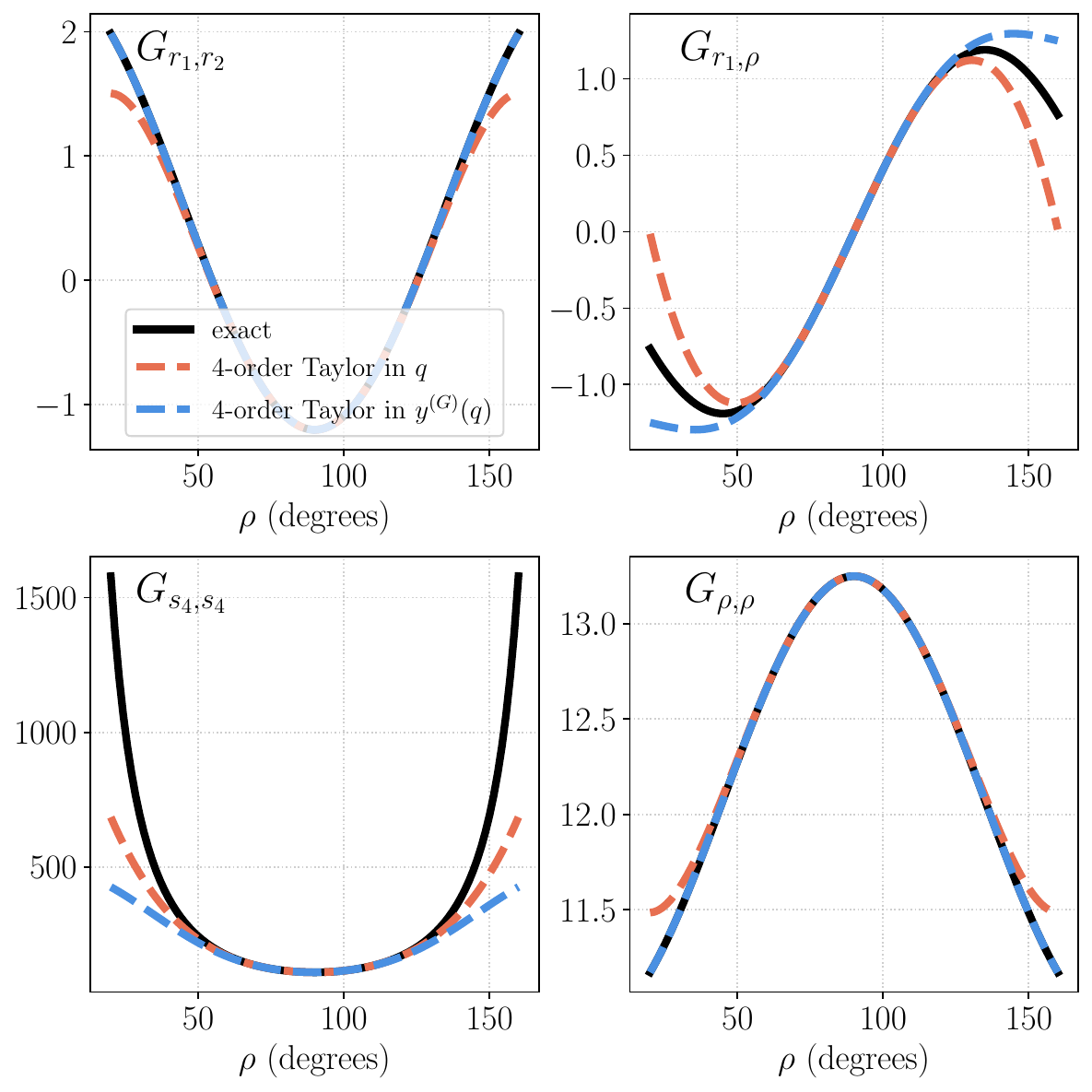}
	\caption{One-dimensional slices of selected elements of the $G$-matrix (in cm$^{-1}$) for NH$_3$, shown along the umbrella coordinate $\rho$.
  Exact values are compared with fourth-order Taylor series expansions around $\rho=90^\circ$, expressed in both original internal coordinates $q=\{r_1,r_2,r_3,s_4,s_5,\rho\}$ and transformed coordinates $y^{(G)}(q)$ in \eqref{eq:y}.}
	\label{fig:nh3_keo}
\end{figure}

\begin{figure}
	\includegraphics[width=\linewidth]{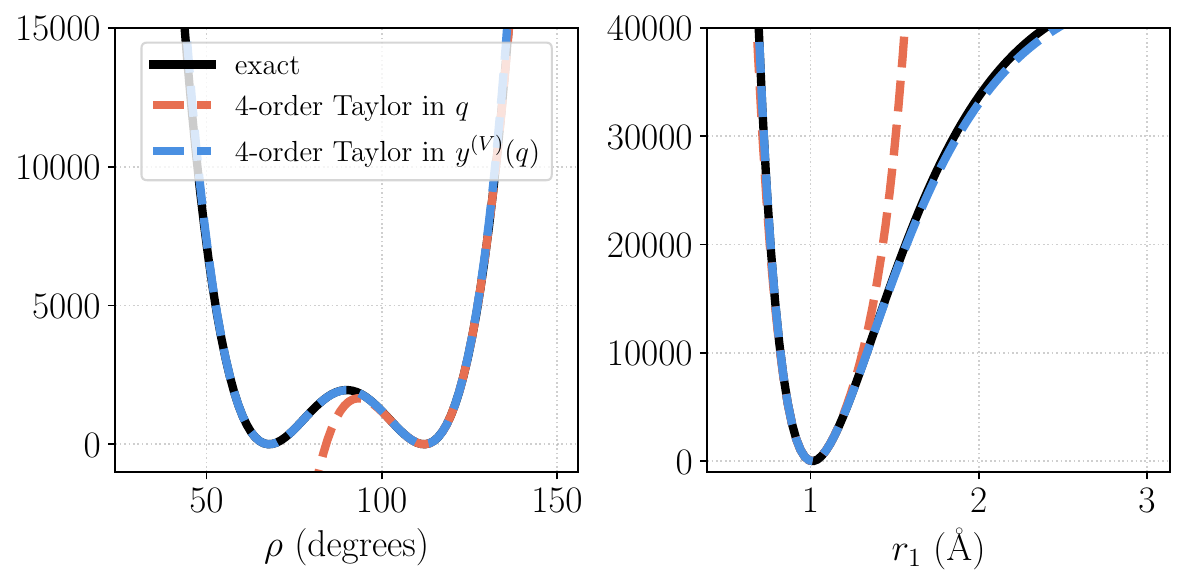}
	\caption{One-dimensional slices of the PES (in cm$^{-1}$) for NH$_3$, shown along the umbrella coordinate $\rho$ and stretching coordinate $r_1$.
  Exact values are compared with fourth-order Taylor series expansions around equilibrium $\rho=112.1^\circ$ and $r_1=1.0116$~\AA, expressed in both internal coordinates $q=\{r_1,r_2,r_3,s_4,s_5,\rho\}$ and transformed coordinates $y^{(V)}(q)$ in \eqref{eq:y}.}
	\label{fig:nh3_pes}
\end{figure}

To demonstrate the efficiency of the Taylor expansion in the transformed $y^{(G)}$ and $y^{(V)}$ coordinates, we present in \autoref{fig:nh3_keo} and \autoref{fig:nh3_pes} one-dimensional slices of selected elements of the $G$-matrix and PES for NH$_3$ along the umbrella coordinate $\rho$.
These slices, calculated exactly on a grid, are compared with fourth-order Taylor expansions in both the transformed coordinates and the original internal coordinates.
The results show that expansions in $y^{(G)}$ and $y^{(V)}$ coordinates yield significantly higher accuracy than those in the original internal coordinates at the same truncation order.
This improvement is especially notable for the PES, where the expansion in internal coordinates around one of the minima fails to reproduce the characteristic double-well feature.

To solve the vibrational problem, we implemented a basis set contraction procedure that involves several steps of reduced-mode variational calculations (see, e.g.,~\cite{Yurchenko_JPCA113_2009}).
In the first step, we defined a primitive one-dimensional basis set for each coordinate and computed matrix elements of the coordinate powers and momentum operators (i.e., $x^t$ and $x^t\partial/\partial x$) appearing in the Taylor expansions of the KEO and PES.
We used Hermite functions as primitive basis for all coordinates of both molecules.
In the second step, one-dimensional (1D) reduced-mode Schrödinger equations were solved independently for each coordinate. Each equation used the primitive basis for the target coordinate with the Hamiltonian averaged over the remaining coordinates using the zero-order primitive basis functions.
The resulting eigenfunctions were then truncated using an energy threshold (typically 40\,000--60\,000~cm$^{-1}$), producing six sets of 1D contracted basis functions, one for each coordinate.

Next, we constructed a combined basis for the equivalent stretching coordinates: $r_2$ and $r_3$ for H$_2$CO and $r_1$, $r_2$, and $r_3$ for NH$_3$.
This was done by solving reduced-mode Schrödinger equations using the previously obtained 1D contracted basis functions for the stretching coordinates.
As in the 1D case, the Hamiltonian was averaged over the remaining coordinates using the ground-state contracted basis functions.
The same procedure was applied in parallel to the bending coordinates, i.e., $\theta_1$ and $\theta_2$ for H$_2$CO, and $s_4$ and $s_5$ for NH$_3$.
The resulting eigenfunctions were again truncated using an energy threshold (30\,000~cm$^{-1}$) and used to form new contracted basis sets for stretching and bending vibrations.

In the final step, the full vibrational basis was constructed as a direct product of contracted basis functions for all coordinates.
To reduce the basis set size, we included only the product functions whose total energy falls below a specified threshold.
The total energy is estimated as a sum of the individual energies of the contracted basis functions.
The resulting full-dimensional vibrational Schrödinger equation was then solved to obtain the vibrational energies and wavefunctions.
Since our goal is to demonstrate the convergence behavior of the KEO and PES Taylor expansions, we employed relatively modest basis sets by setting the energy threshold for the product basis at 12\,000~cm$^{-1}$ for both molecules.

\begin{figure}
	\includegraphics[width=\linewidth]{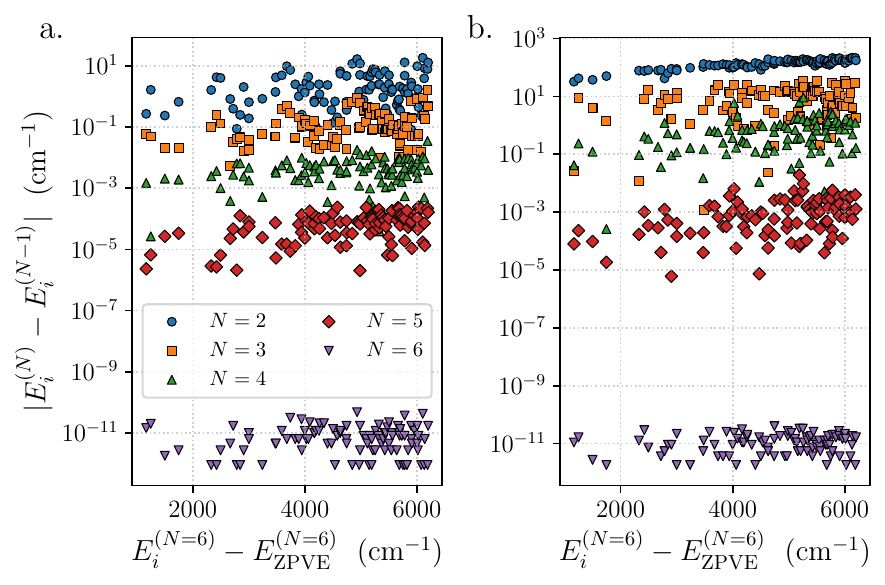}
	\caption{Convergence of the KEO and PES $N$-mode expansions for the first 100 vibrational energy levels of H$_2$CO.
  Plotted are the absolute differences between vibrational energies $E_i$ ($i=1..100$) computed using $N$ and $N-1$ expansion orders, for $N=2..6$, shown as functions of the corresponding vibrational energies relative to the zero-point energy.
  Panel (a) shows convergence for the $G$-matrix and pseudopotential, and panel (b) for the PES.
  }\label{fig:h2co_nmode}
\end{figure}

\begin{figure}
	\includegraphics[width=\linewidth]{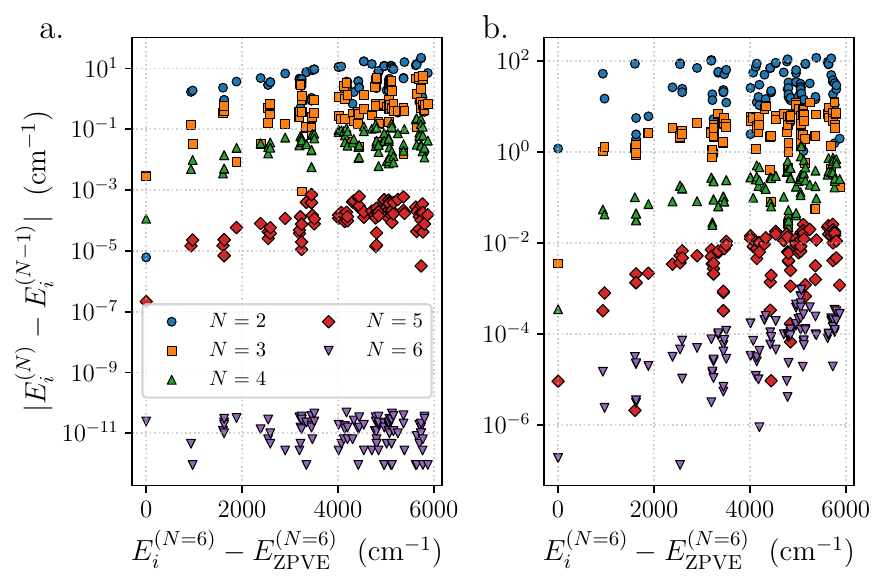}
	\caption{Convergence of the KEO and PES $N$-mode expansions for the first 100 vibrational energy levels of NH$_3$.
  Plotted are the absolute differences between vibrational energies $E_i$ ($i=1..100$) computed using $N$ and $N-1$ expansion orders, for $N=2..6$, shown as functions of the corresponding vibrational energies relative to the zero-point energy.
  Panel (a) shows convergence for the $G$-matrix and pseudopotential, and panel (b) for the PES.
  }\label{fig:nh3_nmode}
\end{figure}

In \autoref{fig:h2co_nmode} and \autoref{fig:nh3_nmode}, we show the convergence of the first 100 vibrational energy levels of H$_2$CO and NH$_3$ with respect to the truncation order of the $N$-mode expansion for the kinetic $G$-matrix and pseudopotential (panel a) and the PES (panel b).
The results are presented as absolute energy differences between calculations performed with $N$ and $N-1$ truncation levels, plotted against the energy values relative to the zero-point energy.
In all calculations, the maximum order of the Taylor series expansions for both the KEO and PES was fixed at 8th order.
When testing the $N$-mode convergence of the $G$-matrix and pseudopotential, the PES expansion was held fixed at 8th order, and vice versa.
The results for both molecules demonstrate fast convergence of the $N$-mode expansion for the KEO operator, and a bit slower convergence for the PES.
Overall, a truncation at $N=4$ is sufficient to achieve sub-wavenumber accuracy for the lowest 100 vibrational states.

\begin{figure}
	\includegraphics[width=\linewidth]{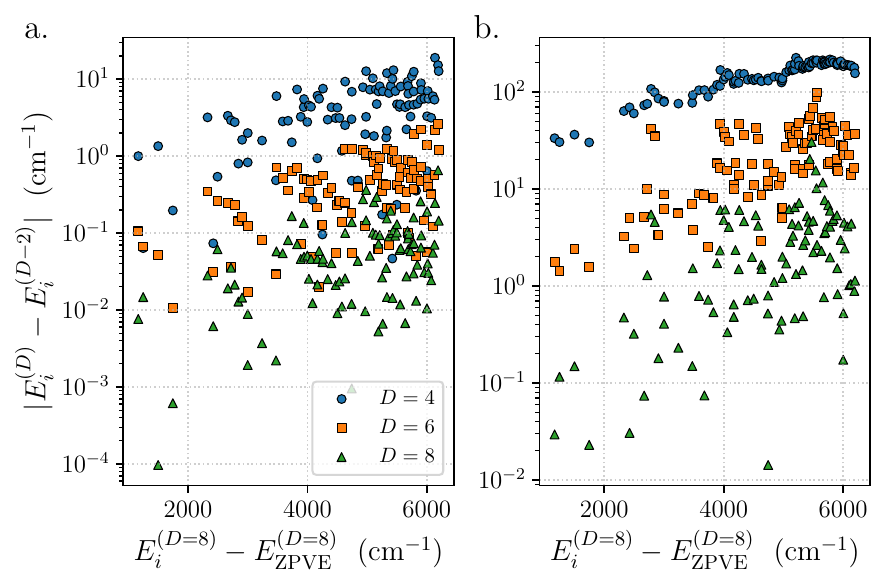}
	\caption{Convergence of the KEO and PES Taylor series expansions for the first 100 vibrational energy levels of H$_2$CO.
  Plotted are the absolute differences between vibrational energies $E_i$ ($i=1..100$) computed using Taylor truncation orders $D$ and $D-2$, for $D=4,6,8$, shown as functions of the corresponding vibrational energies relative to the zero-point energy.
  Panel (a) shows convergence for the $G$-matrix and pseudopotential, and panel (b) for the PES.
  }\label{fig:h2co_pow}
\end{figure}

\begin{figure}
	\includegraphics[width=\linewidth]{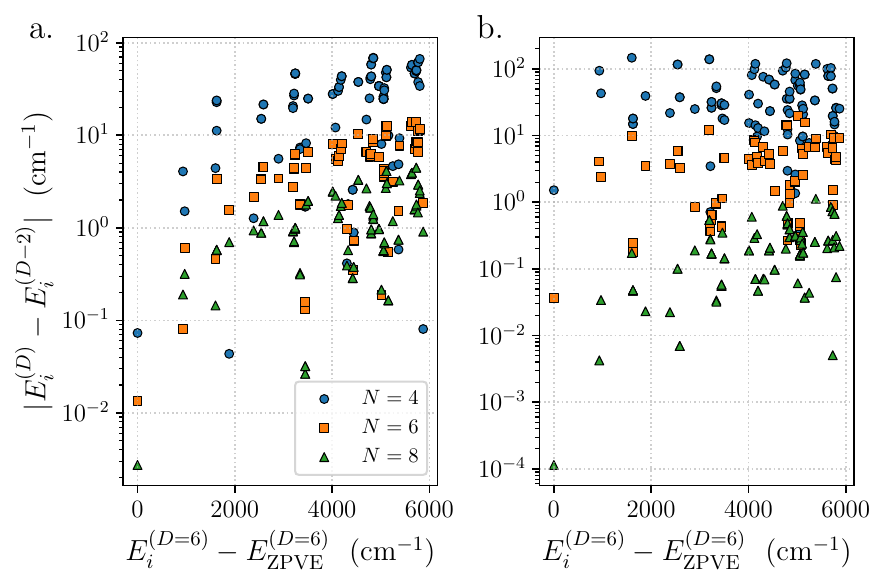}
	\caption{Convergence of the KEO and PES Taylor series expansions for the first 100 vibrational energy levels of NH$_3$.
  Plotted are the absolute differences between vibrational energies $E_i$ ($i=1..100$) computed using Taylor truncation orders $D$ and $D-2$, for $D=4,6,8$, shown as functions of the corresponding vibrational energies relative to the zero-point energy.
  Panel (a) shows convergence for the $G$-matrix and pseudopotential, and panel (b) for the PES.
  }\label{fig:nh3_pow}
\end{figure}

The convergence of vibrational energies for H$_2$CO and NH$_3$ with respect to the truncation order of the Taylor series expansion is shown in \autoref{fig:h2co_pow} and \autoref{fig:nh3_pow}, respectively.
The plots present the absolute energy differences between calculations using Taylor expansions of orders $D$ and $D-2$, plotted against the corresponding energy values relative to the zero-point energy.
For H$_2$CO, convergence of the KEO is achieved within 1~cm$^{-1}$ at the 6th expansion order, with the difference between $D=8$ and $D=6$ being less than 1~cm$^{-1}$.
For NH$_3$, the KEO converges more slowly, which we attribute to the choice of bending coordinates $\beta_{ij}$ (see \autoref{fig:nh3_h2co_coo}.b), where higher-order terms are necessary for accurate expansions.
The PES expansion for NH$_3$ converges faster than for H$_2$CO, due to the use of transformed coordinates $y^{(V)}$ in \eqref{eq:y}, particularly the use of Morse functions for stretching coordinates.

\section{Conclusions}

We presented \texttt{Vibrojet}, a Python implementation of a general framework for constructing rovibrational KEOs and PESs for arbitrary molecules, using user-defined internal coordinate systems and frame embedding conditions.
The framework supports efficient evaluation of rovibrational operators either on grids of internal coordinates or as truncated Taylor series expansions.

The implementation leverages Taylor-mode automatic differentiation capabilities from the \texttt{JAX} library, specifically its \texttt{jet} module, and includes extensions tailored to the specific needs associated with rovibrational operator construction.
These include, for example, efficient Taylor expansion of the KEO in the Eckart frame embedding.

We demonstrated the utility of the framework through variational calculations of molecular vibrational energies based on Taylor ($N$-mode) expansions of the KEO and PES.
The results confirm good convergence behavior of the computed energies with respect to the expansion order.

The present approach based on Taylor series expansion can be combined with traditional least-squares fitting methods to balance the accuracy and the computational cost of high-order expansions.
For example, Taylor expansions can be performed at multiple reference points, and fitting approaches can then be used to construct a unified representation by interpolating the corresponding derivatives.

\section{Data availability}
The data supporting the findings of this study are available within the article and through the following repository: \url{https://github.com/robochimps/vibrojet}

\section{Code availability}
The \texttt{Vibrojet} package developed in this work is available at: \url{https://github.com/robochimps/vibrojet}

\section{Acknowledgements}
We thank Benoit Richard for proofreading the manuscript and for his helpful suggestions.
The work of EV was supported by the European Union's Horizon Europe research and innovation programme under Marie Skłodowska-Curie grant agreement No. 101155136.
The work of AFC was supported by the Data Science in Hamburg HELMHOLTZ Graduate School for the Structure
of Matter (DASHH, HIDSS-0002).

\bibliography{AD, theory_rovibrations, math, PES}%
\onecolumngrid%
\listofnotes%
\end{document}